# Origin of Contact Resistance at Ferromagnetic Metal-Graphene Interfaces

Khoong Hong Khoo, [∥, 1,2,3] Wei Sun Leong, [∥, 2,4] John T. L. Thong, [*, 2,4]

and Su Ying Quek, [*, 1,2]

[1] Department of Physics, National University of Singapore, 2 Science Drive 3, Singapore 117551.

[2] Centre for Advanced 2D Materials and Graphene Research Centre, National University of Singapore, 6 Science Drive 2, Singapore 117542.

[3] Institute of High Performance Computing, 1 Fusionopolis Way, #16-16 Connexis, Singapore 138632.

[4] Department of Electrical and Computer Engineering, National University of Singapore, Singapore 117583.

[∥] These authors contributed equally to this work.

*E-mail: phyqsy@nus.edu.sg (theory); elettl@nus.edu.sg (experiment)

ABSTRACT

Edge contact geometries are thought to yield ultralow contact resistances in most non-ferromagnetic metal-graphene interfaces owing to their large metal-graphene coupling strengths. Here, we examine the contact resistance of edge- versus surface-contacted ferromagnetic metal-graphene interfaces (i.e. nickel- and cobalt-graphene interfaces) using both single-layer and few-layer graphene. Good qualitative agreement is obtained between theory and experiment. In particular, in both theory and experiment, we observe that the contact resistance of edge-contacted ferromagnetic metal-graphene interfaces is much lower than that of surface-contacted ones, for all devices studied and especially for the single-layer graphene systems. We show that this difference in resistance is not due to differences in the metal-graphene coupling strength, which we quantify using Hamiltonian matrix elements. Instead, the larger contact resistance in surface contacts results from spin filtering at the interface, in contrast to the edge-contacted case where both spins are transmitted. Temperature-dependent resistance measurements beyond the Curie temperature $T_C$ show that the spin degree of freedom is indeed important for the experimentally measured contact resistance. These results show that it is possible to induce a large change in contact resistance by changing the temperature in the vicinity of $T_C$, thus paving the way for temperature-controlled switches based on spin.

There has been great interest in using graphene as on-chip interconnects and high-speed transistors in next generation electronic circuits due to its high electron mobility, large thermal conductivity, and highly tunable carrier concentration.[1, 2, 3] Graphene has also been considered a promising replacement for the more expensive indium tin oxide in flexible electronics applications due to its high transmittance, flexibility, and low sheet resistance.[4, 5, 6] On the other hand, the small atomic number of carbon gives rise to a weak spin-orbit interaction and long spin coherence lengths in graphene, making graphene an ideal candidate for spintronics.[7, 8, 9, 10] Graphene is also used as an interconnect between bulk metallic leads and other 2D materials.[11, 12, 13, 14] However, proper control of contact resistance ($R_C$) at the metal-graphene interfaces remains elusive.[15, 16] Metal oxide semiconductor field-effect transistors (MOSFETs) at the current technology node call for a contact resistance of 80 $\Omega.\mu m$ or less per contact,[17] much smaller than typical $R_C$ values measured for graphene devices that range from hundreds up to $10^6$ $\Omega.\mu m$.[16, 18, 19, 20] Furthermore, control of contact resistance at the metal-graphene interfaces is critical for the development of graphene spintronics,[9, 11] as impedance matching is generally required for efficient spin injection rates.

Numerous techniques have been employed to improve the contact resistance of graphene devices. Some, including the use of plasma treatment,[21, 22] UV ozone treatment,[23] and thermal annealing,[18, 24] are based on the premise that the removal of graphene surface contaminants improve the contact interface quality.[25, 26] However, this premise was found invalid for the case of thermal annealing.[27] The use of different contact metals has also been studied both theoretically[28, 29, 30, 31] and experimentally[16, 19, [32,33], where it was found that physisorbed metals give higher contact resistance than chemisorbed metals.

Another thread of investigations that has picked up recently stems from the realization of "edge-contacted", as opposed to "surface-contacted" metal-graphene geometries, where the graphene edge is in direct contact with the metal.[11, 34] Calculations have shown that edge contacts lead to lower contact resistances due to stronger metal-graphene coupling strength.[35, 36] Such edge-contacts have been fabricated by making lithographically defined cuts in the contact region of graphene,[37] encapsulating graphene between boron nitride sheets and etching to make a 1D edge which is then metallized,[36] creating well-defined etched pits in the contact region of graphene,[38] or controlled plasma processing.[39] The "edge contact" geometries have relatively small contact resistances, with the lowest values reported to date being 100 Ω.μm for exfoliated single layer graphene,[36, 38] and 23 Ω.μm for highly n-doped CVD graphene.[40] It was thus deduced that the graphene-metal coupling at edge contacts is stronger, leading to a lower contact resistance in these measurements.

In this paper, we bring attention to the significant role of electron spin in governing the contact resistance of ferromagnetic metal-graphene interfaces. We report a comprehensive study on the contact resistance of ferromagnetic metal-graphene interfaces for both types of contact geometries (*i.e.* edge- and surface-contacted). Good qualitative agreement is obtained between theory and experiment. In particular, we found that the contact resistance of edge-contacted interfaces is indeed much lower than that of the corresponding surface-contacted interfaces. This difference in contact resistance between edge- and surface-contacted interfaces is largest for single layer graphene systems, but persisted also for the few-layer graphene case. Unlike previous studies that attributed the lower edge contact resistance to a difference in metal-graphene coupling, our first-principles calculations show that the higher contact resistance of the surface-contacted Ni-graphene interface arises predominantly from a strong spin filtering effect. This spin filtering effect arises from a majority spin transmission gap near the Fermi level for the graphene surface on Ni, due to conservation of crystal momentum in the surface Brillouin Zone for electrons transmitting between graphene and Ni. For edge contacts, this conservation requirement is relaxed, and no transmission gap is present. Through a temperature-dependence study, we further observe that the contact resistance of edge-contacted Ni-graphene interface shows a dramatic increase once Ni loses its ferromagnetism, showing that the spin degree of freedom is indeed important for the experimentally measured contact resistance. These results show that it is possible to induce a large change in contact resistance by changing the temperature in the vicinity of the Curie temperature $T_C$, thus paving the way for temperature-controlled switches based on spin.

To understand the origin of contact resistance at the ferromagnetic metal-graphene interfaces, we performed first-principles density functional theory (DFT) transport calculations using the Atomistix toolkit package that employs the Non-Equilibrium Green's Function method. We constructed both surface- and edge-contacted Ni-graphene interfaces for single layer graphene (SLG), bilayer graphene (BLG), and trilayer graphene (TLG) systems as shown in Figures 1a-c, as well as surface- and edge-contacted Co-SLG interfaces. Few-layer graphene (FLG) was considered as it is known to give lower contact resistances,[32] reduced sensitivity to ambient conditions, and greater noise tolerance relative to SLG. The surface-contact geometry shown here was chosen to ensure that the contact region is entirely in the scattering region. We have also performed calculations where the surface contact region is repeated infinitely as part of our electrode (Figure 1a; full-surface-contact). Zigzag graphene edges with no hydrogen termination were selected for edge contacts as these are predominantly produced by our contact treatment process[38] (see Supporting Information S1 and S2 for details). Further calculation details are given in "Methods".

Figure 1d plots the transmission spectra for our calculated metal-graphene systems. It can be seen that the transmission increases with the number of graphene layers and is higher for edge-contacted interfaces relative to surface-contacted ones. We also see that the close to the Fermi level, transmission for the SLG surface and edge contacts looks very similar for Co and Ni contact metals. Furthermore, the transmission of surface contacted Ni-SLG interfaces is relatively independent of whether the Ni-graphene overlap is contained entirely within the scattering region ("surface contact") or included as part of the electrode ("full surface contact"). This is consistent with previous calculations and experiments that show the resistance of surface contacted Ni-graphene interfaces to be relatively independent of contact area.[19, 41] The thermally-averaged contact resistance ($R_C$) can be extracted from the transmission using:

$$\frac{1}{R_C} = \int \frac{T(E)}{L} \frac{e^{(E-(E_F+eV_g))/k_BT}}{\left(1+e^{(E-(E_F+eV_g))/k_BT}\right)^2} \frac{dE}{k_BT} \qquad (1)$$

where $T(E)$ is the transmission spectrum, $L$ is the unit cell length along the periodic x-direction, $E_F$ the Fermi level, $E$ the energy, $V_g$ the gate voltage and $T$ the temperature. According to equation (1), $R_C$ is largely determined by transmission values within $k_BT$ of ($E_F + eV_g$). The resulting contact resistance values for $T = 300$ K in the absence of gate voltage are shown in Figure 1e. As can be seen, the trends for the contact resistance are consistent with those of transmission, with smaller values of contact resistance for edg-contacts and larger number of graphene layers.

The carrier concentration in graphene can be readily tuned by the application of a back-gate voltage.[1] Our calculations show that $R_C$ decreases with increasing magnitudes of $V_g$ as the carrier concentration is increased when one moves away from the Dirac point (Figure 2f below). Furthermore, $R_C$ for edge-contacted interfaces is consistently lower than that of corresponding surface-contacted interfaces for the $V_g$ values studied (-0.3 V to 0.3 V). In particular, we observe that $R_C$ of edge-contacted interfaces can be several times smaller than that of surface-contacted interfaces at finite gate voltages, with $R_C$ for edge and surface contacts being 24 Ω.μm and 106 Ω.μm, respectively, when $V_g = 0.3$ V.

Using CMOS-compatible processes, we fabricated and compared SLG and FLG graphene field-effect transistors (FETs) with different types of source/drain contacts, namely untreated (surface-contacted) and treated (edge-contacted) Ni-graphene contacts, as illustrated in Figure 2a-b (see Figure S1 and Methods for detailed fabrication procedures). For all transistors, electrical measurements were conducted in vacuum at room temperature and no annealing was performed prior to measurements. A four-point probe technique was used to extract the contact resistance for each graphene FET, with

$$R_C = \frac{1}{2}(R_{2p} - R_{4p})W \qquad (2)$$

where $R_C$ is the contact resistance, $R_{2p}$ is the device's two-point resistance, $R_{4p}$ is the device's four-point resistance and $W$ is the contact width as indicated in Figure S1d. This definition of contact resistance is not quantitatively the same as that in the theoretical predictions. Nevertheless, good qualitative agreement is obtained between theory and experiment, as shown in Figures 2c-f.

Previously, Yu *et al*. had shown that a back gate bias of 10 V relative to the Fermi level shifts the energy levels of a typical long graphene channel by about 0.1 V relative to the $E_F$ of the electrodes.[42] This explains the choice of theoretical and experimental gate voltages in Figure 2. Similar to the above theoretical predictions, we find experimentally that the contact resistance of Ni-graphene systems: (1) is smaller whenever edge contacts are present, with the largest difference observed for SLG, and regardless of the gate voltage considered here, (2) decreases as the number of graphene layers increases, and (3) decreases with increasing magnitude of $V_g$.

The trends for $R_C$ as a function of the number of graphene layers can be understood from the larger density of states available for transmission near $E_F$ in FLG than in SLG, due to a flattening of bands in the former (Figure S3). This effect is partially compensated in the edge-contacted geometry due to much larger transmission coefficients for SLG, explaining the smaller drop in contact resistance going from SLG to FLG for edge-contacts compared to surface-contacts (see Supporting Information S3). By computing the current density for the minority spin states around $E_F$ at the $k$-point with the largest transmission eigenvalue (see SI), we see that only the first two graphene layers nearest to Ni conduct for surface contacts while all graphene layers conduct for edge contacts. However, the relatively weak variation of contact resistance with number of layers for the edge-contacted case, compared to the difference in $R_C$ between surface and edge contacts (especially for SLG), indicates that the larger $R_C$ for surface contacts is *not* predominantly due to the difficulty of electrons hopping across layers in the FLG systems.

It has been shown previously for a series of non-ferromagnetic metal-graphene interfaces that the contact resistance is reduced by the use of edge over surface contacts, and this was largely attributed to the stronger coupling between graphene and the metal surface in the former. A stronger interface coupling that is relevant for transmission would be quantitatively measured by the size of the Hamiltonian matrix elements (hopping terms) between Ni and graphene. Thus, to examine if lower contact resistance at the edge contacts results from stronger "coupling", we computed the Hamiltonian matrix elements between Ni and graphene orbitals at the interface for both surface- and edge-contacted geometries, as shown in Figure 3. It can be seen that the matrix elements between Ni and C for the surface and edge contacted interfaces are similar in magnitude. In fact, the coupling strength for the Ni $d_{yz}$ and $d_{x^2-y^2}$ orbitals and C $p_y$-orbitals is significantly larger for surface-contacted Ni-SLG interface relative to edge-contacted ones. This demonstrates that the coupling at Ni-graphene edge contacts is *not* larger than that at Ni-graphene surface contacts, and thus, the smaller edge contact resistance cannot be attributed to stronger coupling at graphene edges.

Instead, we find that the larger contact resistance for surface contacts results from a lack of majority spin transmission close to $E_F$. In Figure 4a, we plot the spin-resolved transmission spectra for both surface- and edge-contacted Ni-graphene interfaces with SLG, BLG and TLG as well as those for Co-SLG interfaces. It can be seen that the minority spin transmission for both surface- and edge-contacted interfaces show V-shaped transmission profiles that are quite similar in magnitude. This is consistent with the fact that coupling strength is similar for surface and edge contacts, and is not a limiting factor for transmission at surface contacts. For the majority spin transmission, however, only the edge-contacted interfaces show significant transmission while those of surface contacts have clear transmission gaps where the transmission is nearly zero. The transmission gap lowers the overall current and is the main reason for the larger contact resistance in surface-contacted Ni- and Co-graphene interfaces relative to edge-contacted ones, as depicted schematically in Figure 4c.

The spin filtering may also have potential applications in spintronics, such as in magnetoresistance, when current flows between two magnetic electrodes contacting the graphene channel. In this case, an antiparallel configuration, where the spins in the two electrodes are aligned in opposite directions, will result in an even smaller current for surface-contacted geometries. We also note that even in the presence of randomly aligned magnetic domains, our theoretical predictions relating spin filtering with contact resistance are still valid, because the electronic structure involving majority and minority spins is independent of the direction of easy axis for magnetization.

The absence of majority spin transmission in surface-contacted Ni- and Co-graphene interfaces has been observed previously[43] and it can be understood from the bandstructure of Ni-SLG as shown in Figure 4b. In the majority spin bandstructure, we see an energy window about $E_F$ where there is an absence of states with significant weight on graphene, while the corresponding window for minority spin states only begins at ~ 0.25 eV above $E_F$. This gives rise to a transmission gap near $E_F$ only for majority but not minority spin carriers, as seen in Figure 4a.

Let us also discuss why conductance gaps are only observed for surface graphene contacts but not edge contacted ones. Consider a graphene sheet adsorbed on a Ni (111) surface with the surface normal pointing in the *y*-direction (e.g. Figure 1a). Conservation of crystal momentum requires that $k_x$ and $k_z$ be conserved for electrons transmitting between Ni and graphene. However, the Dirac cone structure of graphene states implies only a very limited volume of the Ni Brillouin zone (BZ) can participate in conduction near $E_F$. This greatly increases the chance of transmission gaps forming. Conversely, for edge contacted Ni-graphene interfaces as shown in Figure 1, there is only periodicity in the *x*-direction and only $k_x$ is conserved. This increases the proportion of Ni BZ available for conduction and the chance of a transmission gap forming is greatly reduced.

In the above calculations, we have shown that the larger contact resistance measured for the surface contacts compared to the edge contacts are likely to arise from spin filtering in the surface contacts. This conclusion is strengthened by previous experimental demonstrations of spin injection in Co-graphene surface contacts.[44] To further confirm this, we performed measurements on the surface and edge contacts over a range of temperatures above and below the Curie temperature ($T_C$) of nickel. As can be seen from Figure 5, the $R_C$ of both surface- and edge-contacted Ni-graphene interfaces show a definite increase around $T_C$ (~ 627 K), when Ni becomes paramagnetic; this is especially so for the case of edge-contacted Ni-graphene where the increase in $R_C$ around $T_C$ is rather dramatic. While these results do not provide direct evidence for our prediction that spin filtering decreases the contact resistance for surface-contacted geometries, the results do demonstrate clearly that the spin degree of freedom in nickel plays an important role in determining the $R_C$ of Ni-graphene interfaces. We approximate the effects of paramagnetism in the Ni electrodes using non-spin-polarized DFT conductance calculations. Although such calculations typically do not describe the paramagnetic state of ferromagnets such as Fe and Co, self-consistent disordered local moment calculations have shown that a virtual-crystal like approximation of the different spins is adequate for describing paramagnetic bulk Ni (see SI).[45] Our calculations show that the transmission near $E_F$ for both surface- and edge-contacted Ni-graphene interfaces decrease when going from spin-polarized to non-spin-polarized results, while this decrease is especially significant for the edge-contacted case for energies below the Dirac point (Figure S5). These results suggest that the experimental observations are due to changes in the electronic structure of Ni in the ferromagnetic-paramagnetic transition. A more detailed discussion is presented in the Supporting Information S5. More sophisticated calculation methodologies[45] may provide a better understanding of $R_C$ above $T_C$, but are beyond the scope of our current work.

In summary, we have presented a detailed investigation of the contact resistance of ferromagnetic Ni- and Co-graphene interfaces for both types of contact geometries (*i.e.* edge- and surface-contacted). We observe that the contact resistance of edge-contacted Ni- and Co-graphene interfaces is much lower than that of surface-contacted interfaces, especially for the single-layer graphene systems. These results are corroborated by good qualitative agreement with experimental studies of Ni-graphene interfaces using both surface and edge contacts. By computing the Hamiltonian matrix elements, we show that the larger contact resistance at the surface-contacted Ni- and Co-graphene interface is not due to differences in metal-graphene coupling strength. Instead, we show that the larger contact resistance at surface contacts results from a strong spin filtering effect that arises from the presence of a majority spin gap near the Fermi level. Due to a relaxation of the *k*-conservation requirement at edge contacts, no spin gap exists for edge contacts. We demonstrate through a temperature dependence study that the spin state of Ni has a profound impact on the contact resistance of the Ni-graphene interface. Our work suggests that one can have a better control of the contact resistance in graphene devices (especially for spintronic applications) by tuning the spin state of magnetic metals.

**Methods**

   **Calculation details.** DFT calculations were performed using the Atomistix toolkit package which employs the Non-Equilibrium Green's Function method for simulating open boundary conditions.[46] The electron density is expanded in a double-$\zeta$ polarization pseudoatomic basis set and a real space mesh cutoff of 150 Ry was used for our calculations. An in-plane k-point mesh perpendicular to the transport direction with spacing 0.17 Å$^{-1}$ was used to obtain the charge density while a much denser *k*-point mesh with spacing 0.0051 Å$^{-1}$ was used for obtaining the transmission spectrum. The exchange-correlation potential is obtained within the local spin density approximation (LSDA).[47] Structural relaxation was carried out using the SIESTA program[48] and performed until all force components on the atoms are less than 0.05 eV/Å.

   The close-packed Ni(111) and Co(0001) surfaces are chosen because of the predominance of close-packed surface facets in experiment and the excellent lattice match with graphene. We found that the most energetically favourable binding sites for graphene edge atoms on Ni(111) are as follows: hollow sites for SLG, one C on a hollow site and one on a top site for BLG, and for TLG, the C atom in the middle graphene layer is on the top site while those of adjacent graphene layers are on hollow sites. In all three cases, the Ni-C bond lengths are less than 2 Å, indicative of strong covalent bonding.

For surface-contacted and full-surface-contacted Ni-graphene systems (see Figure 1), the carbon atoms rest on the top and hollow sites of the Ni surface. In the former, the graphene sheet rests on an elevated island of Ni to effectively isolate the graphene edge from Ni, which constitutes the electrode on one side. For the full-surface-contact system, graphene on Ni constitutes the electrode. The other electrode is graphene. We find the relaxed Ni-C distances to be less than 2.2 Å, indicative of strong covalent binding consistent with previous findings.[31]

For the Co-graphene interfaces, we start with the relaxed geometries of surface- and edge-contacted Ni-SLG systems and re-optimize the geometry using the same convergence criterion as that used for the Ni calculations. This choice of starting geometry is justified by the fact that Co and Ni contacts have similar crystal structures, with nearest-neighbour distances within 1% of each other, and also common binding characteristics to graphene.[43] We find that the bond lengths change by less than 0.1 Å after relaxation.

**Fabrication of graphene FETs with surface-contacted and zigzag-edge-contacted Ni-graphene electrodes.** Graphene flakes were exfoliated on an oxidized degenerately p-doped silicon substrate with 285 nm thick silicon dioxide. The sample was then spin-coated with a 200 nm thick layer of polymethylmethacrylate (PMMA) 950 A4 (Microchem Inc.) and baked at 180 °C on a hot plate for 3 min. Each graphene flake was delineated into a 2 µm wide strip using electron beam lithography (EBL) followed by oxygen plasma etching (20 W RF power, 80 V substrate bias, for 30 s). The sample was then dipped in acetone for 12 h to remove the PMMA layer. The next step is the formation of source/drain contacts. For all FETs with surface-contacted Ni-graphene electrodes, the source/drain contacts were delineated using EBL and metallized with 80 nm of Ni.

On the other hand, for all FETs with edge-contacted Ni-graphene electrodes, the source/drain contacts were delineated using EBL followed by a light oxygen plasma treatment (5 W RF power, 9 V substrate bias, for 10 s), which creates defects in graphene (or few-layer graphene) at the source/drain regions only, while the channel region remains intact as it was protected by the PMMA layer (Figure S1a). A thin Ni film (2 nm thick) was then deposited at the source/drain regions followed by a lift-off process using acetone (Figure S1b). Subsequently, the sample was annealed at 580 °C for 5 min in a hydrogen-rich environment; specifically, the chamber was filled with a 1:2 mixture of hydrogen and argon at a total gas flow rate of 200 sccm at a pressure of 20 Torr. This 3-step contact treatment (*i.e.,* light oxygen plasma, thin Ni film deposition and thermal annealing) is able to yield multiple through-hole etched pits in graphene (or FLG up to 5 layers thick) enclosed by well-defined zigzag edges based on a Ni-catalyzed gasification process: C (solid) + $2H_2$ (gas) → $CH_4$ (gas) (Figure S1c).[38] We have shown previously that a similar procedure results in only zigzag edges in graphene.[38] For verification purposes, Raman analysis and AFM studies were also conducted (see Supporting Information Figure S2 for more details). It should be noted that in the event of light oxygen plasma being omitted from the contact treatment, the etching process most likely takes place in the uppermost graphene layer only, while the bottom graphene layers of FLG remain as perfect $sp^2$-hybridized carbon layers, even for bilayer graphene, as discussed in our earlier work.[13] Finally, the source/drain contacts were delineated and metallized with 80 nm of Ni (Figure S1d).

The dimensions for all graphene FETs in this work are the same. The graphene channel width, channel length, contact width and contact length for all FETs are 2 μm.

## ASSOCIATED CONTENT

**Supporting Information**

Fabrication process of graphene FETs with treated Ni-graphene electrodes, characterization of treated graphene, DFT calculations and layer dependence, contact area dependence of $R_c$ for surface contacted Ni-SLG, non-spin-polarized calculations. This material is available free of charge *via* the Internet at http://pubs.acs.org.

## AUTHOR INFORMATION

**Corresponding Author**

*E-mail: phyqsy@nus.edu.sg (theory); elettl@nus.edu.sg (experiment)

**Notes**

The authors declare no competing financial interest.

## ACKNOWLEDGMENTS

K.H.K. and S.Y.Q. acknowledge support from the Singapore National Research Foundation under grant NRF-NRFF2013-07. Computations were performed on the NUS Graphene Research Centre cluster. We acknowledge support from the Singapore National Research Foundation, Prime Minister's Office, under its medium-sized centre program. W.S.L. and J.T.L.T. acknowledge support from grant R-263-000-A76-750 from the Faculty of Engineering, NUS.

**References**

1. Novoselov KS, *et al.* Electric field effect in atomically thin carbon films. *Science* **306**, 666-669 (2004).

2. Balandin AA, *et al.* Superior thermal conductivity of single-layer graphene. *Nano Letters* **8**, 902-907 (2008).

3. Wu Y, *et al.* State-of-the-Art Graphene High-Frequency Electronics. *Nano Letters* **12**, 3062-3067 (2012).

4. Kim KS, *et al.* Large-scale pattern growth of graphene films for stretchable transparent electrodes. *Nature* **457**, 706-710 (2009).

5. El-Kady MF, Strong V, Dubin S, Kaner RB. Laser scribing of high-performance and flexible graphene-based electrochemical capacitors. *Science* **335**, 1326-1330 (2012).

6. Georgiou T, *et al.* Vertical field-effect transistor based on graphene-WS2 heterostructures for flexible and transparent electronics. *Nature Nanotechnology* **8**, 100-103 (2013).

7. Han W, Kawakami RK, Gmitra M, Fabian J. Graphene spintronics. *Nature Nanotechnology* **9**, 794-807 (2014).

8. Guimaraes MHD, Zomer PJ, Ingla-Aynes J, Brant JC, Tombros N, van Wees BJ. Controlling Spin Relaxation in Hexagonal BN-Encapsulated Graphene with a Transverse Electric Field. *Physical Review Letters* **113**, 5 (2014).

9. Wang Y, *et al.* Room-Temperature Ferromagnetism of Graphene. *Nano Letters* **9**, 220-224 (2009).

10. Young AF, *et al.* Spin and valley quantum Hall ferromagnetism in graphene. *Nature Physics* **8**, 550-556 (2012).

11. Allain A, Kang J, Banerjee K, Kis A. Electrical contacts to two-dimensional semiconductors. *Nature Mater* **14**, 1195-1205 (2015).

12. Avsar A, *et al.* Air-stable transport in graphene-contacted, fully encapsulated ultrathin black phosphorus-based field-effect transistors. *ACS Nano* **9**, 4138-4145 (2015).

13. Leong WS, Luo X, Li Y, Khoo KH, Quek SY, Thong JTL. Low Resistance Metal Contacts to MoS2 Devices with Nickel-Etched-Graphene Electrodes. *ACS Nano* **9**, 869-877 (2015).

14. Wu Y, *et al.* Negative compressibility in graphene-terminated black phosphorus heterostructures. *Physical Review B* **93**, 035455 (2016).

15. Novoselov KS, Fal'ko VI, Colombo L, Gellert PR, Schwab MG, Kim K. A roadmap for graphene. *Nature* **490**, 192-200 (2012).

16. Xia FN, Perebeinos V, Lin YM, Wu YQ, Avouris P. The origins and limits of metal-graphene junction resistance. *Nature Nanotechnology* **6**, 179-184 (2011).

17. The International Technology Roadmap for Semiconductors, Semiconductor Industry Association, 2013.

18. Balci O, Kocabas C. Rapid thermal annealing of graphene-metal contact. *Applied Physics Letters* **101**, 243105 (2012).

19. Nagashio K, Nishimura T, Kita K, Toriumi A. Contact resistivity and current flow path at metal/graphene contact. *Applied Physics Letters* **97**, 143514 (2010).

20. Malec CE, Elkus B, Davidović D. Vacuum-annealed Cu contacts for graphene electronics. *Solid State Communications* **151**, 1791-1793 (2011).

21. Robinson JA, *et al.* Contacting graphene. *Applied Physics Letters* **98**, 053103 (2011).

22. Chu T, Chen ZH. Understanding the Electrical Impact of Edge Contacts in Few-Layer Graphene. *ACS Nano* **8**, 3584-3589 (2014).

23. Wei Chen C, *et al.* UV ozone treatment for improving contact resistance on graphene. *Journal of Vacuum Science and Technology B* **30**, 060604 (2012).

24. Cheng Z, Zhou Q, Wang C, Li Q, Wang C, Fang Y. Toward Intrinsic Graphene Surfaces: A Systematic Study on Thermal Annealing and Wet-Chemical Treatment of SiO2-Supported Graphene Devices. *Nano Letters* **11**, 767-771 (2011).

25. Lin Y-C, Lu C-C, Yeh C-H, Jin C, Suenaga K, Chiu P-W. Graphene annealing: how clean can it be? *Nano Letters* **12**, 414-419 (2011).

26. Pirkle A, *et al.* The effect of chemical residues on the physical and electrical properties of chemical vapor deposited graphene transferred to SiO2. *Applied Physics Letters* **99**, 122108 (2011).

27. Leong WS, Nai CT, Thong JTL. What Does Annealing Do to Metal–Graphene Contacts? *Nano Letters* **14**, 3840-3847 (2014).

28. Barraza-Lopez S, Vanević M, Kindermann M, Chou MY. Effects of Metallic Contacts on Electron Transport through Graphene. *Physical Review Letters* **104**, 076807 (2010).

29. Gong C, Lee G, Shan B, Vogel EM, Wallace RM, Cho K. First-principles study of metal–graphene interfaces. *Journal of Applied Physics* **108**, 123711 (2010).

30. Ran Q, Gao M, Guan X, Wang Y, Yu Z. First-principles investigation on bonding formation and electronic structure of metal-graphene contacts. *Applied Physics Letters* **94**, 103511 (2009).

31. Khomyakov PA, Giovannetti G, Rusu PC, Brocks G, van den Brink J, Kelly PJ. First-principles study of the interaction and charge transfer between graphene and metals. *Physical Review B* **79**, 195425 (2009).

32. Venugopal A, Colombo L, Vogel EM. Contact resistance in few and multilayer graphene devices. *Applied Physics Letters* **96**, 013512 (2010).

33. Ji X, Zhang J, Wang Y, Qian H, Yu Z. A theoretical model for metal-graphene contact resistance using a DFT-NEGF method. *Physical Chemistry Chemical Physics* **15**, 17883-17886 (2013).

34. Xu Y*, et al.* Contacts between Two- and Three-Dimensional Materials: Ohmic, Schottky, and p–n Heterojunctions. *ACS Nano* **10**, 4895-4919 (2016).

35. Matsuda Y, Deng WQ, Goddard WA. Contact Resistance for "End-Contacted" Metal-Graphene and Metal-Nanotube Interfaces from Quantum Mechanics. *Journal of Physical Chemistry C* **114**, 17845-17850 (2010).

36. Wang L*, et al.* One-Dimensional Electrical Contact to a Two-Dimensional Material. *Science* **342**, 614-617 (2013).

37. Smith JT, Franklin AD, Farmer DB, Dimitrakopoulos CD. Reducing Contact Resistance in Graphene Devices through Contact Area Patterning. *ACS Nano* **7**, 3661-3667 (2013).

38. Leong WS, Gong H, Thong JTL. Low-Contact-Resistance Graphene Devices with Nickel-Etched-Graphene Contacts. *ACS Nano* **8**, 994-1001 (2014).

39. Yue DW, Ra CH, Liu XC, Lee DY, Yoo WJ. Edge contacts of graphene formed by using a controlled plasma treatment. *Nanoscale* **7**, 825-831 (2015).

40. Park HY*, et al.* Extremely Low Contact Resistance on Graphene through n‐Type Doping and Edge Contact Design. *Advanced Materials*, (2015).

41. Stokbro K, Engelund M, Blom A. Atomic-scale model for the contact resistance of the nickel-graphene interface. *Physical Review B* **85**, 165442 (2012).

42. Yu Y-J, Zhao Y, Ryu S, Brus LE, Kim KS, Kim P. Tuning the Graphene Work Function by Electric Field Effect. *Nano Letters* **9**, 3430-3434 (2009).

43. Maassen J, Ji W, Guo H. Graphene Spintronics: The Role of Ferromagnetic Electrodes. *Nano Letters* **11**, 151-155 (2011).

44. Kamalakar MV, Dankert A, Bergsten J, Ive T, Dash SP. Enhanced Tunnel Spin Injection into Graphene using Chemical Vapor Deposited Hexagonal Boron Nitride. *Scientific Reports* **4**, (2014).

45. Staunton J, Gyorffy BL, Pindor AJ, Stocks GM, Winter H. Electronic-structure of metallic ferromagnets above the Curie-temperature. *Journal of Physics F-Metal Physics* **15**, 1387-1404 (1985).

46. Brandbyge M, Mozos J-L, Ordejón P, Taylor J, Stokbro K. Density-functional method for nonequilibrium electron transport. *Physical Review B* **65**, 165401 (2002).

47. Perdew JP, Zunger A. Self-interaction correction to density-functional approximations for many-electron systems. *Physical Review B* **23**, 5048-5079 (1981).

48. José MS*, et al.* The SIESTA method for ab initio order- N materials simulation. *Journal of Physics: Condensed Matter* **14**, 2745 (2002).

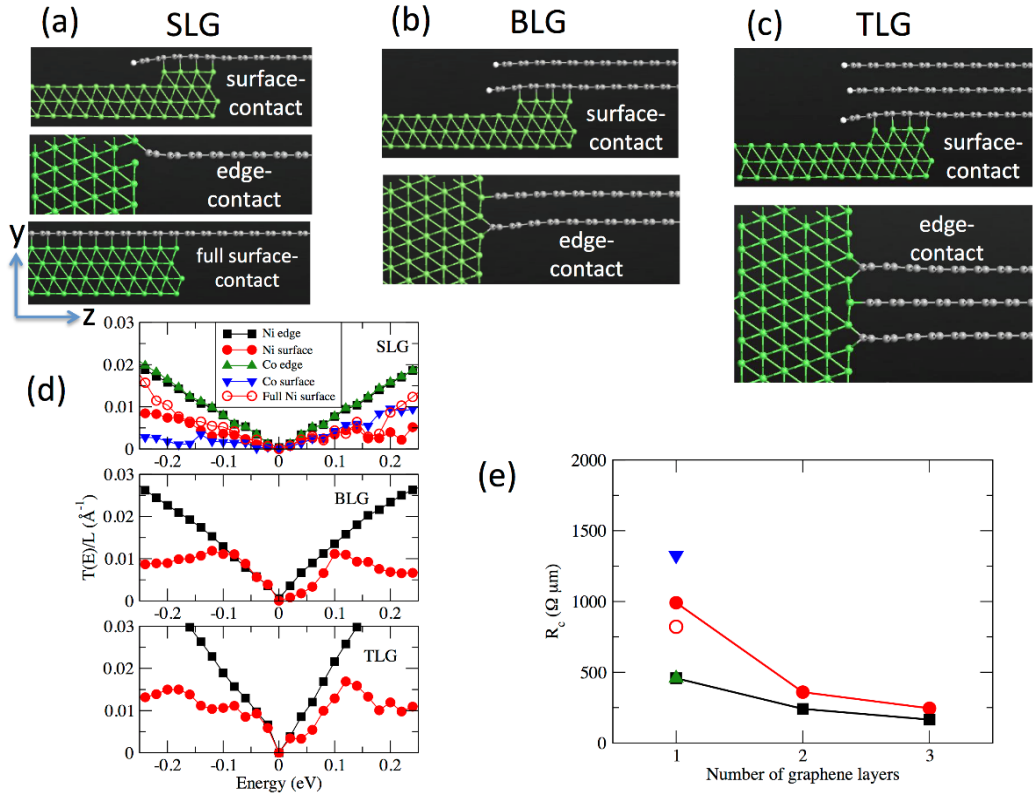

**Figure 1.** (a-c) Optimized atomic geometries of both surface and edge contacted Ni-graphene interfaces for single layer graphene (SLG), bilayer graphene (BLG), and trilayer graphene (TLG) systems employed in our DFT transport calculations. In each case, SLG, BLG or TLG constitutes the other electrode, corresponding to semi-infinite graphene sheets. "Full surface-contact" refers to the case where graphene on Ni is part of the left electrode. (d) Transmission coefficient per transverse line segment at zero bias for the edge- and surface-contacted Ni- and Co-graphene interfaces. (e) Contact resistance versus number of graphene layers extracted when $V_g = 0$ V. The legend is indicated in (d).

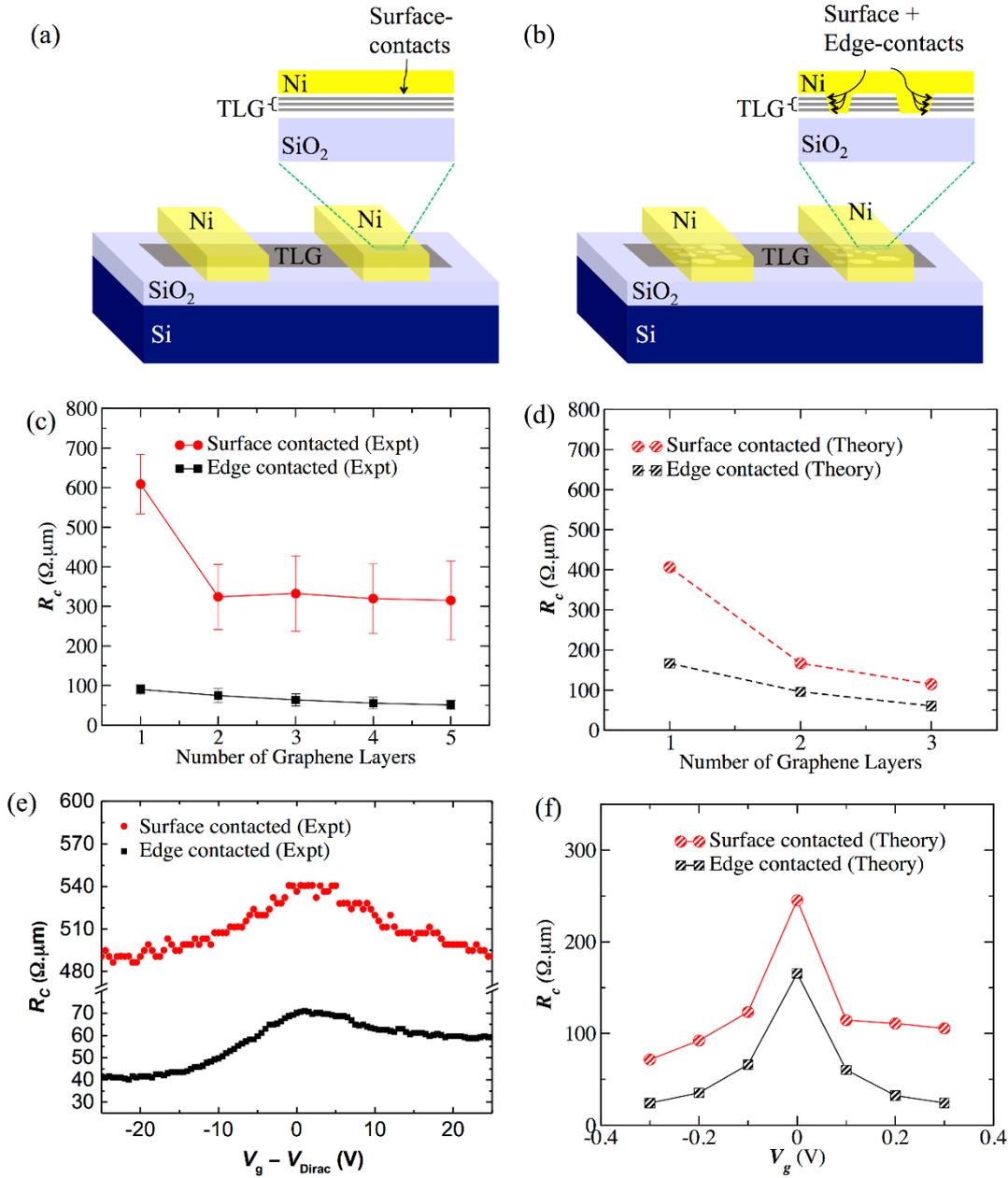

**Figure 2.** (a) Schematic of a back-gated trilayer graphene (TLG) FET with untreated (*i.e.* surface-contacted) Ni-graphene electrodes. (b) Schematic of a back-gated TLG FET with treated (*i.e.* edge-contacted) Ni-graphene electrodes. (c) Experimental contact resistance of both edge and surface contacts as a function of the number of graphene layers. Each experimental data point was extracted from at least 10 graphene transistors when $V_g - V_{Dirac} = 10$ V. (d) Theoretical contact resistance versus number of graphene layers calculated for edge and surface contacts with $V_g = 0.1$ V (corresponding to an experimental gate voltage of ~10 V).[42] (e) Measured nickel–TLG contact resistance at 300 K as a function of back-gate voltage. (f) Calculated nickel–TLG contact resistance at 300 K as a function $V_g$ of using equation (1).

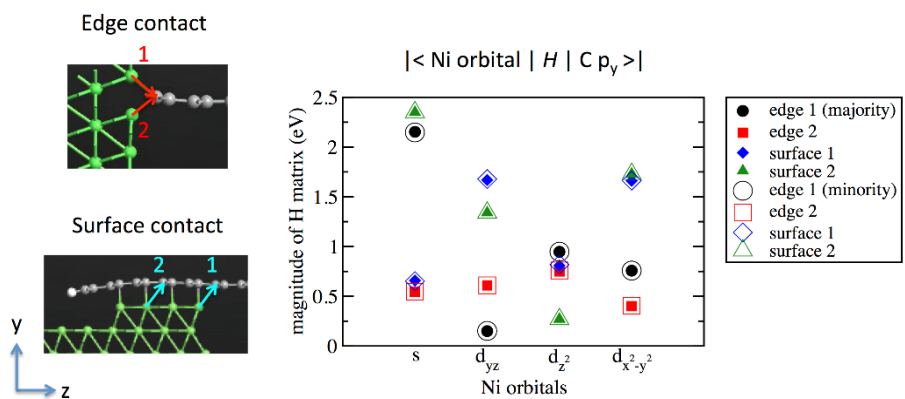

**Figure 3.** Magnitude of majority and minority-spin Hamiltonian matrix elements calculated between graphene-$p_y$ and Ni orbitals centered on atoms connected by arrows in the ball and stick diagrams.

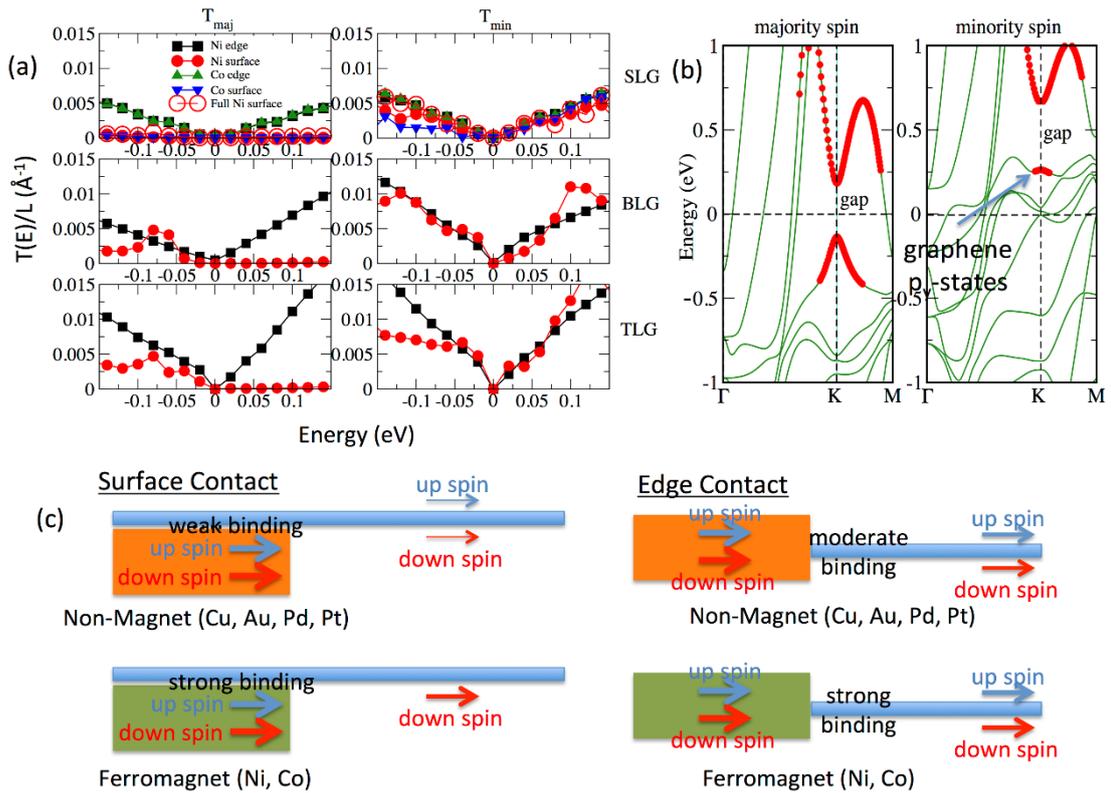

**Figure 4**. (a) Majority (left) and minority (right) spin transmission coefficient per transverse line segment for surface- and edge-contacted Ni- and Co-graphene interfaces. (b) Majority and minority spin bands for Ni-SLG slab (same geometry as in "full-surface-contact" electrode). Red dots represent states where over 50% of wavefunction weight is on graphene $p_y$ orbitals. (c) Schematic of electron flow in edge- and surface-contacted graphene interfaces with magnetic and non-magnetic metal contacts.

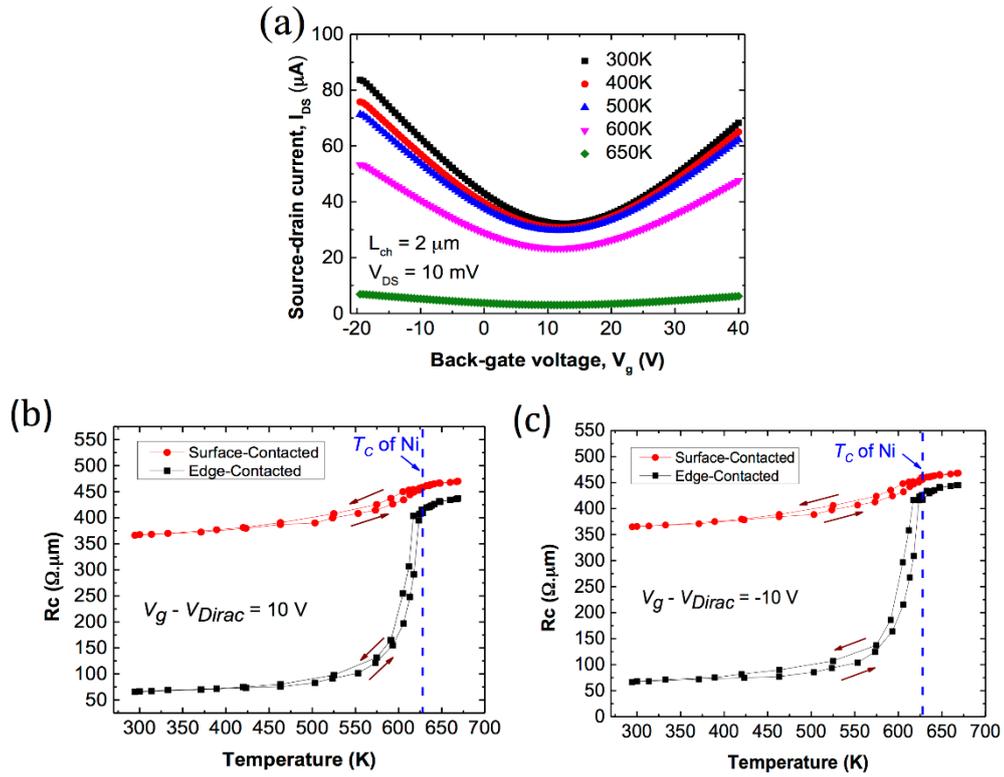

**Figure 5.** (a) Transfer characteristics of a typical edge-contacted trilayer graphene FET at temperatures ranging from 300 to 650 K. (b-c) Measured nickel–graphene contact resistance as a function of temperature at two different gate biases as indicated. Notably, for both cases, the $R_C$ of edge-contacted Ni-graphene interfaces show a dramatic increase around nickel's Curie temperature, $T_C$ (~ 627 K), when Ni becomes paramagnetic.